\documentclass[journal]{IEEEtran}

\usepackage{cite}
\DeclareFontFamily{OT1}{pzc}{}
\DeclareFontShape{OT1}{pzc}{m}{it}{<-> s * [1.10] pzcmi7t}{}
\DeclareMathAlphabet{\mathpzc}{OT1}{pzc}{m}{it}

\usepackage{xcolor}
\usepackage{multirow}
\usepackage{amsbsy,amsfonts,amsfonts,amssymb,epsfig}
\usepackage{graphicx}
\usepackage{graphics}

\usepackage{flushend}

\usepackage{setspace}
\usepackage{color, colortbl}
\definecolor{Gray}{gray}{0.9}
\definecolor{LightCyan}{rgb}{0.88,1,1}
\usepackage[first=0,last=9]{lcg}

\usepackage{color, colortbl}
\definecolor{Gray}{gray}{0.9}
\definecolor{LightCyan}{rgb}{0.88,1,1}
\usepackage[first=0,last=9]{lcg}

\usepackage{color}

\ifCLASSINFOpdf
\else
\fi

\usepackage{amsmath}
\usepackage{textcomp}
\usepackage{mathtools}
\usepackage{enumerate}

\usepackage{color}

\usepackage{xcolor}
\usepackage{multirow}
\usepackage{amsbsy,amsfonts,amsfonts,amssymb,amsmath,epsfig}
\usepackage{graphicx}
\usepackage{graphics}
\usepackage[cmintegrals]{newtxmath}

\makeatletter
\def\vhrulefill#1{\leavevmode\leaders\hrule\@height#1\hfill \kern\z@}
\makeatother

\makeatletter 
\let\myorg@bibitem\bibitem
\def\bibitem#1#2\par{%
	\@ifundefined{bibitem@#1}{%
		\myorg@bibitem{#1}#2\par
	}{%
		\begingroup
		\color{\csname bibitem@#1\endcsname}%
		\myorg@bibitem{#1}#2\par
		\endgroup
	}%
}

\makeatother

\begin{document}

\title{Double Media-Based Modulation Scheme for High-Rate Wireless Communication Systems}
%

\author{Burak Ahmet Ozden,~Erdogan~Aydin,~and Fatih Cogen
	
	\thanks{B. A. Ozden is with the Department of Electrical and Electronics Engineering, Istanbul Medeniyet University, 34857 Istanbul, Turkey, and also with the Department of Computer Engineering, Yildiz Technical University, 34220 Istanbul, Turkey (e-mail: bozden@yildiz.edu.tr).}
	
	\thanks{E. Aydin is with Istanbul Medeniyet University, Department of Electrical and Electronics Engineering, 34857, Uskudar, Istanbul, Turkey (e-mail: erdogan.aydin@medeniyet.edu.tr) (Corresponding author: Erdogan Aydin.)}
	\thanks{F. Cogen is with the 6GEN Laboratory, Next-Generation R\&D, Network Technologies, 34854 Istanbul, Turkey (e-mail: fatih.cogen@turkcell.com.tr).}

 \thanks{This work was supported by The Scientific and Technological Research Council of Turkey (TUBITAK) through the 1515 Frontier Research and Development Laboratories Support Program under Project 5229901 - 6GEN. Lab: 6G and Artificial Intelligence Laboratory. Also, this work was supported by TUBITAK 1001 (Grant Number: 123E513).}

}

\maketitle

\begin{abstract}
Current wireless communication technologies are insufficient in the face of ever-increasing demands. Therefore, novel and high-performance communication systems are needed. In this paper, a novel high data rate and high-performance index modulation scheme called double media-based modulation (DMBM) is proposed. The DMBM system doubles the number of mirror activation patterns (MAPs) and the number of transmitted symbols compared to the traditional MBM system during the same symbol period. In this way, the spectral efficiency of the DMBM is doubled and the error performance improves as the number of bits carried in the indices increases. Performance analysis of the DMBM scheme is evaluated for $M$-ary quadrature amplitude modulation ($M$-QAM) on Rayleigh fading channels. The error performance of the proposed DMBM system is compared with spatial modulation (SM), quadrature SM (QSM), MBM, and double SM (DSM) techniques. Also, the throughput,  complexity, energy efficiency, spectral efficiency, and capacity analyses for the proposed DMBM system and SM, QSM, MBM, and DSM systems are presented. All analysis results show that the proposed DMBM system is superior to the compared systems.

\end{abstract}

\begin{IEEEkeywords}
Media-based modulation, spatial modulation, index modulation, wireless communication, capacity, complexity.
\end{IEEEkeywords}

%
\IEEEpeerreviewmaketitle

\section{Introduction}

\IEEEPARstart{T}{he} need for wireless communication systems with low energy, high spectral efficiency, high energy efficiency, better error performance, and high data-rate has reached its peak today.  multiple-input and multiple-output (MIMO) systems increase the capacity of the system, data rate, and provide better error performance with spatial diversity and multiplexing techniques \cite{Aydin2016}. Novel systems that provide innovative and serious advantages developed integrated into the MIMO system are promising in bringing wireless communication to a new era.

Index modulation (IM) techniques integrated into multiple-input and multiple-output (MIMO) systems aiming to meet these demands and needs continue to gain popularity and importance. IM has brought wireless communication to a new dimension by transmitting information in the indices of the elements of a MIMO system. IM techniques provide lower energy consumption, high spectral efficiency, high data-rate, high-performance, and low complexity compared to traditional techniques \cite{Basarnextg2017, Maoindex2019, Aydin2016,  GCIM, tvt_anten}. A new scheme called orthogonal-frequency division multiplexing (OFDM) with all index modulation  is proposed, which simplifies the system structure by replacing the phase-shift keying (PSK) or quadrature amplitude modulation (QAM) constellation modulator with a subblock modulator in \cite{ShiOFDM}. Also, in \cite{ShiGenetic}, a genetic algorithm based OFDM with an all index modulation scheme is proposed where the subblock realizations in the look-up table are considered as chromosomes of the genetic algorithm.

Media-based modulation (MBM) scheme makes it possible to transmit information in the indices of mirror activation patterns (MAPs) by controlling the radiation pattern of MAPs according to the change of on/off state of radio frequency (RF) mirrors near the transmit antenna \cite{mapburak, AydinMBM2021, LiMBM2021, burak}. In this scheme, the far-field radiation pattern of a reconfigurable antenna with PIN diodes is changed according to the on-off state of the available RF mirrors. The changing of the radiation pattern creates different realizations of the wireless channel. In this way, the MBM technique creates its own modulation alphabet \cite{KhandaniMBM2013,BasarMBM2019, burakris }. Some examples of the important advantages provided by the MBM technique: it increases the number of constellation points without increasing the energy, increases the spectral efficiency linearly depending on the number of RF mirrors, and provides a better error performance than its rival spatial modulation (SM) technique in high spectral efficiency. Also, to increase the data rate, a significant amount of transmitting antennas is required in the SM, while in the MBM technique, it is sufficient to keep the antenna number constant and only increase the RF number.

SM technique uses antenna indices for additional data bit transmission and activates only one antenna during each transmission period, thus eliminating inter-channel interference and providing higher spectral and energy efficiency compared to traditional communication systems \cite{mia, miavv, RISSMCIM}. Inspired by the SM technique in \cite{YigitDSM2016}, an innovative method called double SM (DSM) is proposed, which provides a high data rate and high spectral efficiency by doubling the number of active transmit antennas. In this scheme, it has been observed that the DSM has better error performance than other SM schemes such as quadrature SM (QSM). In  \cite{TsvakiDSMBM2021}, a novel system called double spatial media-based modulation is proposed, which is obtained by adding the MBM technique to the DSM scheme. It is observed that the error performance improved with the addition of MBM to the DSM scheme. A system called double generalized spatial modulation (DGSM), which enables the use of more spatial domain by increasing the number of antenna index vectors in an antenna index vector set, is presented in \cite{Zhao2022}.

\begin{figure*}[t]
\centering{\includegraphics[width=1\textwidth]{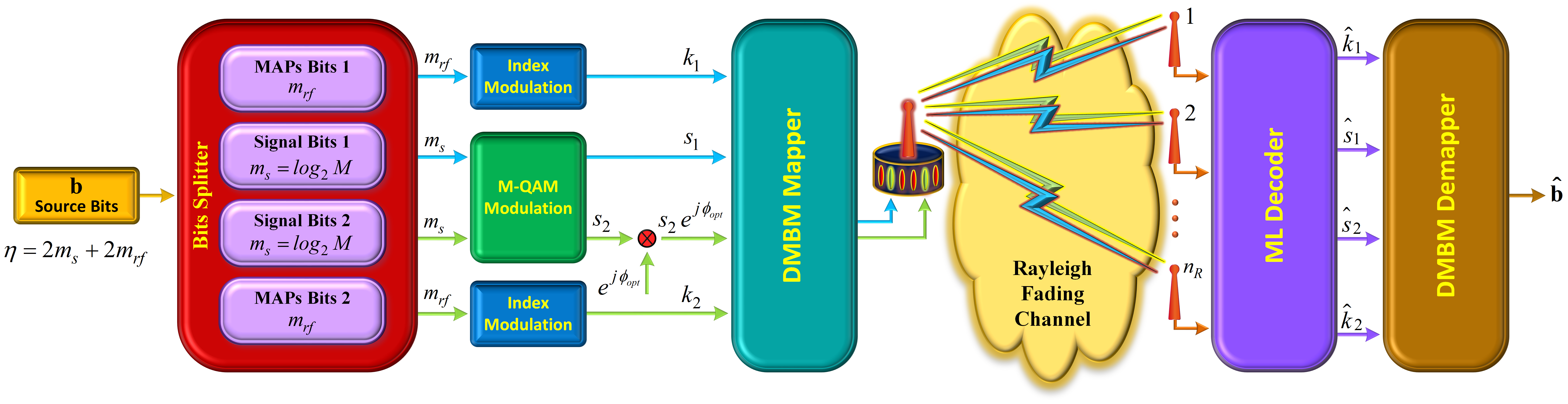}}
\caption{System model of the DMBM system.}
	\label{system_model} 
\end{figure*}



\subsection{Contributions}

In this paper, a novel index modulation scheme called double media-based modulation (DMBM) that doubles the data bits transmitted during a symbol transmission period compared to the traditional MBM system is presented. The proposed DMBM system provides high data-rate, high energy efficiency, and high performance. The contributions of this paper can be summarized as follows:  


\begin{enumerate}

    \item The theoretical average bit error rate (ABER) analyses of the DMBM system are derived and it is shown that the theoretical and simulation results overlap perfectly.  It is shown that the proposed system has better error performance than SM, QSM, MBM, and DSM systems.
    

    \item The capacity analysis of the proposed DMBM system is derived and the results are compared with SM, QSM, MBM, and DSM systems. It is shown that the proposed system has a higher capacity.
    
    \item The computational complexity, throughput, energy efficiency, and spectral efficiency analyses of the proposed DMBM system are obtained and compared with the SM, QSM, MBM, and DSM systems. It is shown that the proposed system is more advantageous in all aspects.

\end{enumerate}

\subsection{Organization and Notation}
The remaining of this paper is organized as follows. Section II includes the system model structure, working principle, and mathematical expression of the proposed system. In Section III, performance analysis including theoretical ABER derivation is presented. In Section IV, complexity, energy efficiency, throughput, and spectral efficiency analysis are given. Section V presents the capacity analyses of the proposed DMBM system. In Section VI, the simulation results are given, which include BER performance comparisons of the theoretical and simulation results of the proposed scheme and benchmark systems. Finally, this paper is concluded with Section VII.

\emph{Notation}: Bold lower/upper case symbols represent vectors/matrices. $(\cdotp)^H$, $(\cdotp)^T$, $\big\|\cdotp \big\| $, $\big| \cdotp \big|$, $E\big\{\cdotp\big\}$, and $Var\big\{ \cdotp \big\}$ denote Hermitian, transpose, Frobenius norm, Euclidean norm, expectation operator, and variance operator, respectively.

\section{DMBM System Model}
The proposed DMBM system model operating over the Rayleigh fading channel is given in Fig. \ref{system_model}. In this system model, it is assumed that the on/off switching time of the RF mirrors creating MAP states is much lower than the symbol transmission time. In Fig. \ref{system_model}, the transmitter is equipped with $m_{rf}$ RF mirrors and one transmit antenna while the receiver is equipped with $n_R$ receive antennas. In the MBM technique, the on/off states of the RF mirrors change according to the information bits. Thus, $2^{m_{rf}}$ different MAPs are obtained to carry information in the indices. This means that $\mathcal{R}=2^{m_{rf}}$ different channel realizations occur to be activated.

\subsection{Principle of DMBM Technique}

In the proposed system model, two MAP states are selected from the available $\mathcal{R}$ MAP states, and two symbols ($s_1$ and $s_2$) are selected from the $M$-QAM symbol constellation. After the optimum rotation angle is applied to  $s_2$, these two symbols are transmitted over two active MAPs. In this way, half of the incoming bits are mapped to the first symbol and first MAPs indices, and the other half to the second symbol and second MAPs indices. In the DMBM scheme, unlike the traditional MBM technique, two symbols and two MAPs are active in the same time interval.  This doubles the data rate (spectral efficiency) of the system compared to the traditional MBM system. Hence, the proposed system transmits  $\eta =2(m_s + m_{rf})$ bits while the MBM scheme conveys $m_s+ m_{rf}$ bits during a transmission period, where $\eta$ is the spectral efficiency in [bits/s/Hz] of the DMBM scheme and $m_s={\log}_2(M)$.

As seen in Fig. \ref{system_model}, firstly, the $\textbf{b}$  bits with the size of $1 \times \eta$ come into the bit splitter. Here, the $\eta$ bits are split into two parts $\textbf{b}_1$ bits with the size of  $1 \times 2m_s$ and $\textbf{b}_2$ bits with the size of $1 \times 2m_{rf}$  so that $\textbf{b}=[\textbf{b}_1 \ \textbf{b}_2]$. Then, half of the $\textbf{b}_1$ bits, i.e., the first $m_s$ bits of  $\textbf{b}_1$ are mapped to the selected first symbol $s_1$, and the other half bits of  $\textbf{b}_1$ select second symbol $s_2 e^{j\phi_{\text{opt}}}$. As seen here, the phase of the selected second symbol is rotated by $\phi_{\text{opt}}$ degree. This angle is the optimum rotation angle for $M$-QAM constellations that minimize the BER of the DMBM scheme. Thus, interference between two symbols is minimized. Similarly, half of the $\textbf{b}_2$ bits i.e., the first $m_{rf}$ bits of  $\textbf{b}_2$ determine the $\textit{k}^{\text{th}}_1$ MAP status of the first RF mirrors and the other half bits of $\textbf{b}_2$ select the second  MAP status $\textit{k}^{\text{th}}_2$ of the second RF mirrors. As a result, $s_1$ is conveyed on the $\textit{k}^{\text{th}}_1$ MAP status and $s_2$ is transmitted on the $\textit{k}^{\text{th}}_2$ MAP status.

\begin{figure}[t]
\centering
\centering{\includegraphics[scale=0.5]{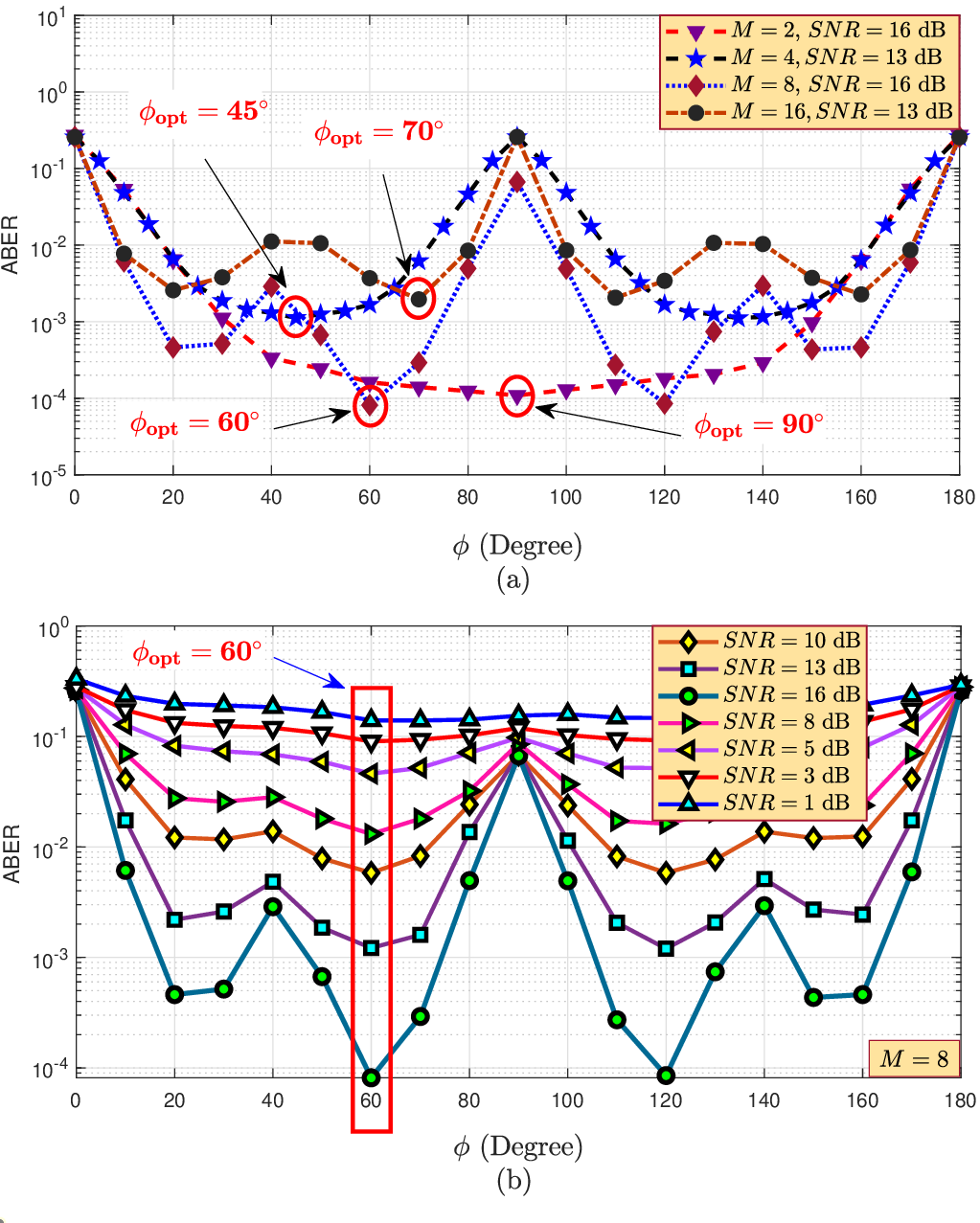}}
	\caption{Rotation angle that minimizes the BER of the DMBM system (a) for $M=2,4,8$ and $16$ at $\text{SNR}=13, 16\ \text{dB}$ while $n_R=4$ and (b) for $M=8$, $n_R=4$ and $ \text{SNR}=1, 3, 5, 8, 10, 13$ \text{and} $16\ \text{dB}$.}
	\label{angle} 
\end{figure}

BER curves corresponding to the rotation angles are given in Fig. \ref{angle}. Here, the rotation angle value corresponding to the minimum BER value is optimum. Also, note that the optimum rotation angle doesn’t change depending on the SNR values as seen in Fig. \ref{angle} (b). Furthermore, it is observed that the optimum angle does not also depend on the wireless fading channel type and its variance. Considering Fig. \ref{angle} (a), the optimum rotation angle $\phi_{\text{opt}}$ is $90^{\circ}$, $45^{\circ}$, $60^{\circ}$, and $70^{\circ}$ for $M=2,4,8,16$ respectively while $\text{SNR}=13, 16\ \text{dB}$ and $m_{rf}=2, n_R=4$.

\subsection{Signal Model and Receiver Structure for DMBM Scheme}

Considering Fig. \ref{system_model}, the transmission vector $\textbf{s}\in \mathbb{C}^{\mathcal{R}\times1 }$ used in the DMBM system can be expressed as:
\begin{eqnarray}\label{transmission_s}
\textbf{s} &=& \Big[ 0 { \ }\cdots { \ } 0 \underbrace{s_1}_{\textit{k}_1^{\text{th}}} { \ } 0 { \ }\cdots { \ } 0  { \ }  \underbrace{s_2 e^{j\phi_{\text{opt}}}}_{\textit{k}_2^{\text{th}}} { \ } 0 { \ }\cdots { \ } 0 \Big]^T,
\end{eqnarray}
where $k_1, k_2 \in \{1, 2, \dots \mathcal{R}\}$ denote the positions of the $s_1$ and $s_2 e^{j\phi_{\text{opt}}}$ symbols. In (\ref{transmission_s}), it can be $k_1=k_2$. However, since the phase of the second symbol is rotated by the degree of $\phi_{\text{opt}}$, the interference of the symbols is minimized.

Consider the case where information bits $\textbf{b}=[10010011]$ ($\textbf{b}_1=[1001]$ and $\textbf{b}_2=[0011]$) will be transmitted at a symbol transmission period for $M=4$ and $m_{rf}=2$. First of all, it should be kept in mind that the optimum shifting angle of $s_2$  should be $\phi_{\text{opt}}=45^{\circ}$ since $M=4$. In this case, the first $2$ bits of $\textbf{b}_1$ ($[10]$) selects the $s_1=1+j$ while the second $2$ bits of $\textbf{b}_1$ ($[01]$) determines the $s_2=-1-j$. After $s_2$ is determined, the rotation angle is applied to $s_2$, i.e., $s_2=(-1-j)e^{j45^{\circ}}= -1.41j$. Similarly, the first and second $2$ bits of $\textbf{b}_2$ and  ($[00],[11]$) determine the $1^{\text{st}}$ and $4^{\text{th}}$ MAPs status of the active RF mirrors, respectively. The transmission vector $\textbf{s}$ for this example is expressed as $\textbf{s} = \big[  1+j  { \ } 0 { \ } { \ } 0  { \ }  -1.41j \  { \ } \big]^T.$


The vector $\textbf{s}$ is transmitted over a Rayleigh fading channel denoted by the channel matrix $\textbf{H} \in \mathbb{C}^{n_R\times \mathcal{R}}$ with mean of zero and variance of $\sigma=1$, i.e., $\sim\mathcal{CN}\big(0,1 \big)$. As a result, the noisy and faded  baseband signal at the receiver is represented as:
\begin{eqnarray} \label{received_signal} \textbf{y} &=& \textbf{Hs} \ + \ \textbf{w}, \nonumber \\ 
&=&  \textbf{h}_{k_1} s_1 \ + \ \textbf{h}_{k_2} s_2 e^{j\phi_{\text{opt}}} \ + \ \textbf{w}, 
\end{eqnarray} 
where $\textbf{y} \in \mathbb{C}^{n_R\times 1}$ is the received signal and $\textbf{w} \in \mathbb{C}^{n_R\times 1}$ is the complex Gaussian noise that has zero mean and variance $\sigma^2=N_0/2$ per dimension, i.e., $\sim\mathcal{CN}(0, N_0)$. $\textbf{h}_{k_1}$ and $\textbf{h}_{k_2}$ are expressed as the $k_1^{\text{th}}$ and $k_2^{\text{th}}$ column vectors of $\textbf{H}$.

Finally, assuming perfect CSI at the receiver, the ML decoder detects active channel states $\hat{k}_1$ and $\hat{k}_2$ of RF mirror 1 and RF mirror 2, also, $\hat{s}_1$ and $\hat{s}_2$ symbols. Consequently, the  ML detector of the DMBM system is obtained as follows:
\begin{eqnarray}\label{ML}
\!\!\!\!\!\!\!\Big[\hat{s}_1,\hat{s}_2,\hat{k}_1,\hat{k}_2\Big] \!\!\!\!& =&\!\!\!\! \text{arg}\underset{s_1,s_2,k_1,k_2}{\mathrm{min}} \bigg|\bigg| \textbf{y} - \Big( \textbf{h}_{k_1} s_1 + \textbf{h}_{k_2} s_2 e^{j\phi_{\text{opt}}}\Big) \bigg|\bigg|^2 .
\end{eqnarray}


\section{Performance Analysis of the DMBM System}

In this section, the ABER of the DMBM system is analyzed.  The original bits of information transmitted from the transmitter may be obtained erroneously at the receiver side, due to the fading effect of the Rayleigh channel and the noise. Examining the pairwise error probability (PEP), which gives information about the probability of making an erroneous decision, is important for theoretical error analysis. When $\textbf{s}$ is transmitted and it is erroneously detected as $\hat{\textbf{s}}$, the conditional PEP (CPEP) of the proposed system is given as follows:
\begin{eqnarray}\label{eq:equationPEP}
\mathcal{P}\Big( \textbf{s} \rightarrow \hat{\textbf{s}} \ \big| \ \textbf{H} \Big)
&=&\mathcal{P}\bigg( \big|\big|\textbf{y} - \textbf{H} \textbf{s} \big|\big|^2 > \big|\big|\textbf{y} - \textbf{H} \hat{\textbf{s}} \big|\big|^2 \bigg).
\end{eqnarray}
The CPEP expression in (\ref{eq:equationPEP}) can be rearranged as follows:
\begin{eqnarray}
        \mathcal{P}\Big( \textbf{s} \rightarrow \hat{\textbf{s}}  \big|  \textbf{H} \Big)  &=&  \mathcal{P}\left(\big\vert\big\vert \mathbf{w}\big\vert\big\vert^2 > \big\vert\big\vert \textbf{H} \textbf{s}+ \mathbf{w}- \textbf{H} \hat{\textbf{s}}  \big\vert\big\vert^2\right)
     \nonumber\\
   &=&  \mathcal{P}\Big( \big|\big| \textbf{H} \textbf{s} \big|\big|^2 -  \big|\big| \textbf{H} \hat{\textbf{s}} \big|\big|^2 \!\! - 2\Re \big\{ \textbf{y}^H\big(\textbf{H} \textbf{s} - \textbf{H} \hat{\textbf{s}}  \big) \big\} \!\!> \!0 \Big) \nonumber  \\
    &=& \mathcal{P} \big( \mathcal{D} > 0 \big),
\label{eq:qEquation_ea}
\end{eqnarray}
where $\mathcal{D} \sim \mathcal{N}(\mu_\mathcal{D},\sigma^2_\mathcal{D})$ is Gaussian distributed R.V, where $\mu_\mathcal{D}$ and $\sigma^2_\mathcal{D}$ can be  given as follows respectively:
\begin{eqnarray}
  \mu_\mathcal{D}  &=&  E\big\{ \mathcal{D} \big\} = E\Big\{-\big|\big| \textbf{H} \textbf{s} -  \textbf{H} \hat{\textbf{s}} \big|\big|^2  - 2\Re \big\{ \textbf{w}^H\big(\textbf{H} \textbf{s} - \textbf{H} \hat{\textbf{s}}  \big) \big\}\Big\} \nonumber\\ 
  &=& - E\Big\{\big|\big| \textbf{H} \textbf{s} -  \textbf{H} \hat{\textbf{s}} \big|\big|^2\Big\} = -\big|\big| \textbf{H} \textbf{s} -  \textbf{H} \hat{\textbf{s}} \big|\big|^2 
\end{eqnarray}
\begin{eqnarray}
  \sigma^2_{\mathcal{D}}  \!\!\!\!&=\!\!\!\!&  Var\big\{ \mathcal{D} \big\} =  Var\Big\{-\big|\big| \textbf{H} \textbf{s} -  \textbf{H} \hat{\textbf{s}} \big|\big|^2  - 2\Re \big\{ \textbf{w}^H\big(\textbf{H} \textbf{s} - \textbf{H} \hat{\textbf{s}}  \big) \big\}\Big\} \nonumber\\ 
  \!\!\!\!&=\!\!\!\!&  4\big|\big|\textbf{H} \textbf{s} - \textbf{H} \hat{\textbf{s}}  \big|\big|^2Var\big( w^H \big)=2N_0\big|\big| \textbf{H} \big( \textbf{s} - \hat{\textbf{s}} \big) \big|\big|^2. 
\end{eqnarray} 
As  known, the $Q(.)$ function is used to calculate the CPEP in (\ref{eq:qEquation_ea}), and the $Q(.)$ function can be defined as $\mathcal{P}\left(Z>z\right) = Q\left(z\right) = \frac{1}{2\pi}\int_{z}^{\infty}\exp\left({\frac{-t^2}{2}}\right)dt$ \cite{simon2005digital},
where $Z \sim \mathcal{N}\left(0,1\right)$.

Hence, considering (\ref{eq:qEquation_ea}), statistical values of $\mathcal{D}$ are re-arranged by normalizing its standard deviation and shifting its mean to zero. Consequently, the CPEP function can be rewritten in terms of $Q(.)$ as follows:
\begin{equation}\label{eq_cpep}
\begin{aligned}
\mathcal{P}\Big( \textbf{s} \rightarrow \hat{\textbf{s}}  \ \big| \  \textbf{H} \Big)   &=Q\left(\frac{-\mu_\mathcal{D}}{\sigma_\mathcal{D}}\right) =Q\Bigg(  \sqrt{\frac{\big|\big| \textbf{H} \big( \textbf{s} - \hat{\textbf{s}}\big) \big|\big|^2}{2N_0}}  \Bigg).   
\end{aligned}
\end{equation}
CPEP in (\ref{eq_cpep}) can be rewritten in integral form  as follows:
\begin{equation}\label{eq3}
\mathcal{P}\Big( \textbf{s} \rightarrow \hat{\textbf{s}} \ \big| \ \textbf{H} \Big) =    \frac{1}{\pi}\int_0^{\pi/2}\exp{\Bigg(\frac{\big|\big| \textbf{H} \big( \textbf{s} - \hat{\textbf{s}} \big) \big|\big|^2}{ 4N_0\sin^2\varphi}\Bigg)}d\varphi.
\end{equation}
Considering a moment generating function (MGF) based approach, the unconditional PEP (UPEP) is obtained by averaging (\ref{eq3}) over the channel matrix \textbf{H}  as follows:
	\begin{equation}\label{eq4}
	\mathcal{P}\Big(\textbf{s} \rightarrow \hat{\textbf{s}}\Big) =    \frac{1}{\pi}\int_0^{\pi/2}\Bigg(\frac{\sin^2{\varphi
}}{\sin^2{\varphi
} + \frac{|| \textbf{s} - \hat{\textbf{s}}  ||^2}{4N_0} }\Bigg)^{n_R}d\varphi.
	\end{equation}
Finally, the closed-form expression of (\ref{eq4}) is derived as \cite{simon2005digital}
\begin{equation}\label{eq5}
\mathcal{P} \Big( \textbf{s} \rightarrow \hat{\textbf{s}} \Big) =  \frac{1}{2} \Bigg[ 1- \mathcal{K} \sum_{k=0}^{n_R-1}{2k \choose k} \Bigg( \frac{1-\mathcal{K}^2}{4} \Bigg) \Bigg],
\end{equation}
where $\mathcal{K}=\frac{||\textbf{s} -\hat{\textbf{s}}||^2}{4N_0+||\textbf{s} - \hat{\textbf{s}}||^2}$. After obtaining the UPEP, by using the well-known upper bounding technique, the theoretical ABER of the DMBM system is described as follows:
\begin{equation} \label{ABER} 
	ABER_{\text{DMBM}} \approx \frac{1}{\eta2^\eta}\sum^{2^\eta}_{i=1}{\sum^{2^\eta}_{j=1}{\mathcal{P}\big( \textbf{s}_i \rightarrow \hat{\textbf{s}}_j \big)\mspace{2mu}\epsilon_{i,j}}}, 
	\end{equation} 
where, $\mathcal{P}\big( \textbf{s}_i \rightarrow \hat{\textbf{s}}_j \big)$ is the average pairwise error probability given in (\ref{eq5}), and $\epsilon_{i,j}$ is defined as the number of bit errors associated with the corresponding pairwise error event.

\definecolor{s0}{rgb}{1, 0.95, 0.7}
\definecolor{s1}{rgb}{1, 0.9, 0.6}
\definecolor{s2}{rgb}{1, 0.85, 0.5}
\definecolor{s3}{rgb}{1, 0.8, 0.4}
\definecolor{s4}{rgb}{1, 0.7, 0.3}
\definecolor{s5}{rgb}{1, 0.65, 0.2}

\begin{table}
\addtolength{\tabcolsep}{-1pt}
  \caption{ Computational Complexity Comparisons for $\eta$ bits}
\centering
\begin{tabular}{|c || c |}

\hline
\rowcolor{s0}
\small System & \small Real Multiplications (RMs)\\

\hline 
   \rowcolor{s1}
     $\mathcal{O}_{\text{SM}}$ \cite{Mesleh2015} &  $8 M n_T n_R \Big( 1 + \frac{m_s+2m_{rf}-m_{sm}}{m_s+m_{sm}} \Big)$
    \\ \rowcolor{s2}
    $\mathcal{O}_{\text{QSM}}$ \cite{Mesleh2015}  & $8 M n_T n_R \Big( 1 + \frac{m_s+2(m_{rf}-m_{sm})}{m_s+2m_{sm}} \Big)$ 
    \\ \rowcolor{s3}
    $\mathcal{O}_{\text{MBM}}$  & $16 M \mathcal{R} n_R$ 
    \\  \rowcolor{s4}
    $\mathcal{O}_{\text{DSM}}$  &  $16 M^2 n^2_T n_R  \Big( 1 + \frac{2(m_{rf}-m_{sm})}{2(m_s+m_{sm})} \Big)$ 
    \\\rowcolor{s5}
  $\mathcal{O}_{\text{DMBM}}$ &  $16 M^2 \mathcal{R}^2 n_R$ 
    \\
\hline

\end{tabular}
\label{ComplexityAnalysis}
\end{table}

\section{Complexity, The Energy Efficiency, Throughput, and Spectral Efficiency Analysis for DMBM System}

In this section, complexity, energy efficiency, throughput, and spectral efficiency analyses, which are important performance criteria, are obtained. Also, the complexity, energy efficiency, throughput, and spectral efficiency analyses of the DMBM scheme are compared with the SM, QSM, MBM, DSM systems.

\subsection{Complexity Analysis}

\begin{figure}[t]
\hspace*{-0.5cm} 
\centering{\includegraphics[scale=0.5]{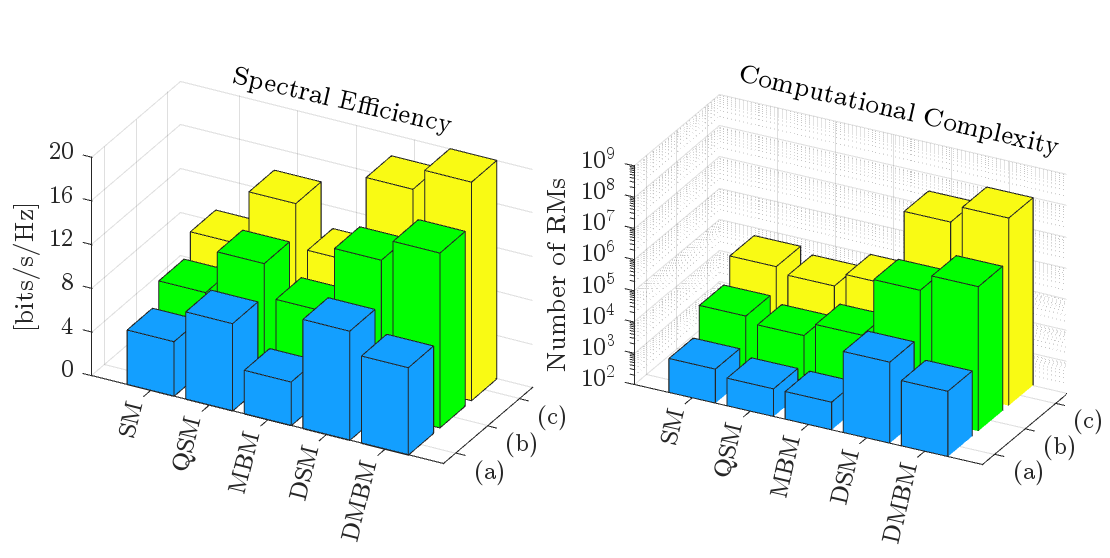}}
	\caption{Spectral efficiency and computational complexity comparisons while $M=4$, $N_T=8$, $m_{rf}=2$, $N_R=3$  for (a); $M=8$, $N_T=16$, $m_{rf}=5$, $N_R=4$  for (b); and $M=16$, $N_T=32$, $m_{rf}=6$, $N_R=6$  for (c).}
	\label{DMBM_comp} 
\end{figure}

In this section, complexity analyses of SM, QSM, MBM,
DSM, and DMBM systems are presented. In Table \ref{ComplexityAnalysis}, computational complexity analysis results are given, where $m_{sm}$ represents the total number of bits transmitted in the transmit antenna indices of the SM system. The overall computational complexities in the receiver terminal of the considered systems are evaluated in terms of real multiplications (RMs). Considering the ML detector (\ref{ML}) of the proposed system, for each of the $2^{\eta}$ decision metrics, $||.||^2$ is performed, and for each $||.||^2$ operation, $2n_R$ RMs are evaluated. In addition, $8$ RMs are performed in the product of the complex channel coefficient and the complex symbol  for each $||.||^2$ operator. As a result, the overall complexity of the DMBM system  can be expressed as $\mathcal{O}_{\text{DMBM}}=16 M^2 \mathcal{R}^2 n_R$. With similar derivations, the computational complexity of the SM system can be calculated as $\mathcal{O}_{\text{SM}}=8 M n_T n_R $ for $\eta_{SM}$ bits. However, when comparing the SM technique with the proposed DMBM system for the same spectral efficiency ($\eta$ bits), the total computational complexity of the SM technique is $\mathcal{O}_{\text{SM}}=8 M n_T n_R \Big( 1 + \frac{m_s+2m_{rf}-m_{sm}}{m_s+m_{sm}} \Big)$. Similarly, when the DMBM scheme and other benchmark systems are compared at the same spectral efficiency ($\eta$ bits), the total computational complexities of the benchmark systems are calculated as in Table \ref{ComplexityAnalysis}. Comparative numerical examples for the computational complexity of SM, QSM, MBM, DSM, and the proposed DMBM systems for different system parameters are presented in Fig. \ref{DMBM_comp}. As can be easily seen from Fig. \ref{DMBM_comp}, when the spectral efficiency of the systems increases, the computational complexity increases as a cost. Also, it is observed that the proposed DMBM system contains a certain amount more computational complexity but provides higher spectral efficiency.

\begin{figure}[t]
\hspace*{-1.5cm} 
\centering{\includegraphics[scale=0.8]{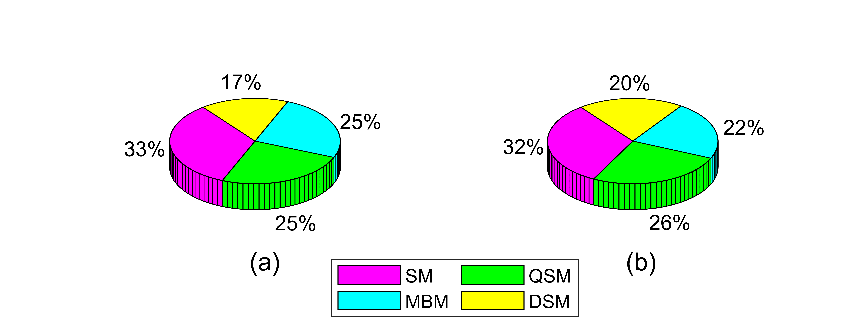}}
	\caption{Energy saving percentages of the proposed DMBM system compared to other compared systems ($M=4$, $n_T=4$, $m_{rf}=4$) for (a), ($M=8$, $n_T=8$, $m_{rf}=8$) for (b).}
	\label{DMBM_energy} 
\end{figure}

\begin{figure}[t]
\hspace*{-0.5cm} 
\centering{\includegraphics[scale=0.58]{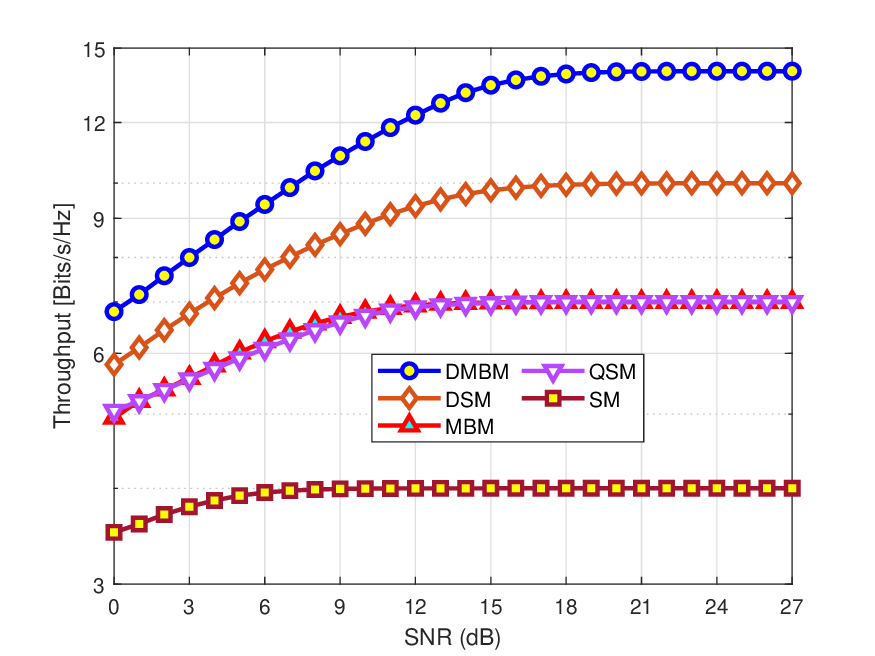}}
	\caption{Throughput of the proposed DMBM, DSM, MBM, QSM, and SM systems ($M=8$, $N_T=4$, $m_{rf}=4$, and $n_R=4$).}
	\label{DMBM_throp} 
\end{figure}


\subsection{Energy Efficiency Analysis}
Nowadays, the importance of energy efficiency is quite obvious considering the high cost caused by the high power consumption of wireless communication systems and environmental problems such as carbon emissions. In the proposed DMBM system, most of the data bits are transmitted at MAP indices. It also provides much more data transmission over a transmission time than traditional techniques. Therefore, the proposed DMBM system provides high energy efficiency. The energy saving percentage $(E_{\text{sav}})$ per $\eta$ bits of the proposed DMBM system compared to SM, QSM, MBM and DSM systems can be defined as follows:
\begin{equation}\label{energy}
E_{\text{sav}}=\Big(1-\frac{\eta_c}{\eta}\Big)E_b\%,
\end{equation}
where $E_b$ is the bit energy and $\eta_c$ represents the spectral efficiency of the compared systems (SM, QSM, MBM, and DSM). The energy efficiency values in percent provided by the proposed DMBM system for different system parameters according to the compared SM, QSM, MBM, and DSM systems are given in Fig. \ref{DMBM_energy}. Also, Fig. \ref{DMBM_energy} shows that the proposed DMBM system provides higher energy efficiency than the other compared systems.


\definecolor{wildwatermelon}{rgb}{1, 0.94, 0.5}
\definecolor{wildwatermelon2}{rgb}{1, 1, 0.65}
\definecolor{pastelpink}{rgb}{1, 1, 0.82}
\begin{table}[t]

\addtolength{\tabcolsep}{-5.6pt}
		\caption{Spectral Efficiency Comparisons  [bits/s/Hz]}
	\begin{center}
		\label{Tablodata}
		\begin{tabular}{|ccc|c|c|c|c|c|} 
		\rowcolor{wildwatermelon}
		   
	        \hline
			\hline
			 &  &   & \small SM & \small QSM & \small MBM & \small DSM & \small DMBM \\ 
	    \rowcolor{wildwatermelon2}
	    	
			\hline
	       \scriptsize $M$ & \scriptsize $n_T$ & \scriptsize $m_{rf}$   & \scriptsize $\log_2(n_TM)$  & \scriptsize $\log_2(n_T^2M)$ & \scriptsize $m_{rf}\!+\!\log_2(M)$ & \scriptsize $\log_2(n_T^2M^2)$ & \scriptsize $2m_{rf}\!+\!\log_2(M^2)$ \\
	    \rowcolor{pastelpink}
			\hline
			\hline
			$4$  &    $2$    & $3$      & $3$   & $4$  & $5$ & $6$ &$10$ \\  
		\rowcolor{pastelpink}
			$8$ &    $4$      & $5$   &    $5$ &   $7$ &    $8$ &    $10$ &$16$  \\   
		\rowcolor{pastelpink}	
			
			$16$  &    $8$      & $7$    &    $7$ &   $10$ &    $11$ &    $14$ & $22$  \\   
			\hline
			\hline
		
		\end{tabular}
	\end{center}
\end{table}



\subsection{Throughput and Spectral Efficiency Analysis}

For a wireless communication system, the throughput can be defined as the amount of data bits that can be accurately obtained at the receiver from the data bits transmitted over a wireless channel. Based on this, the throughput expression of the DMBM system can be defined as follows \cite{Tse}:

\begin{equation}\label{eqtp1}
\mathbb{T}    =  \frac{\big(1-ABER_{\text{DMBM}}\big)}{\uptau _s}\eta,
\end{equation}    
where $\big(1-ABER_{\text{DMBM}}\big)$ represents the probability of bits correctly detected by the receiver from the bits transmitted during the symbol duration $\uptau _s$. Considering that the proposed DMBM system has high spectral efficiency and low error data transmission, it can be easily understood that it is very efficient in terms of throughput. Fig. \ref{DMBM_throp} also shows that the DMBM system provides higher throughput than the other systems.

Spectral efficiency comparisons of the SM, QSM, MBM, DSM, and DMBM systems are presented in Table \ref{Tablodata}. Depending on the increase in $M$, the number of transmit antennas ($n_T$), and $m_{rf}$ values, it is seen that the system with the highest spectral efficiency of increase among the SM, QSM, MBM, DSM, and DMBM systems is the DMBM system. Additionally, spectral efficiency values of SM, QSM, MBM, DSM, and the proposed DMBM systems are visualized for different system parameters in Fig. \ref{DMBM_comp}. It is shown that the proposed DMBM system provides higher spectral efficiency than other systems.




\begin{figure}[t]
\centering{\includegraphics[scale=0.5]{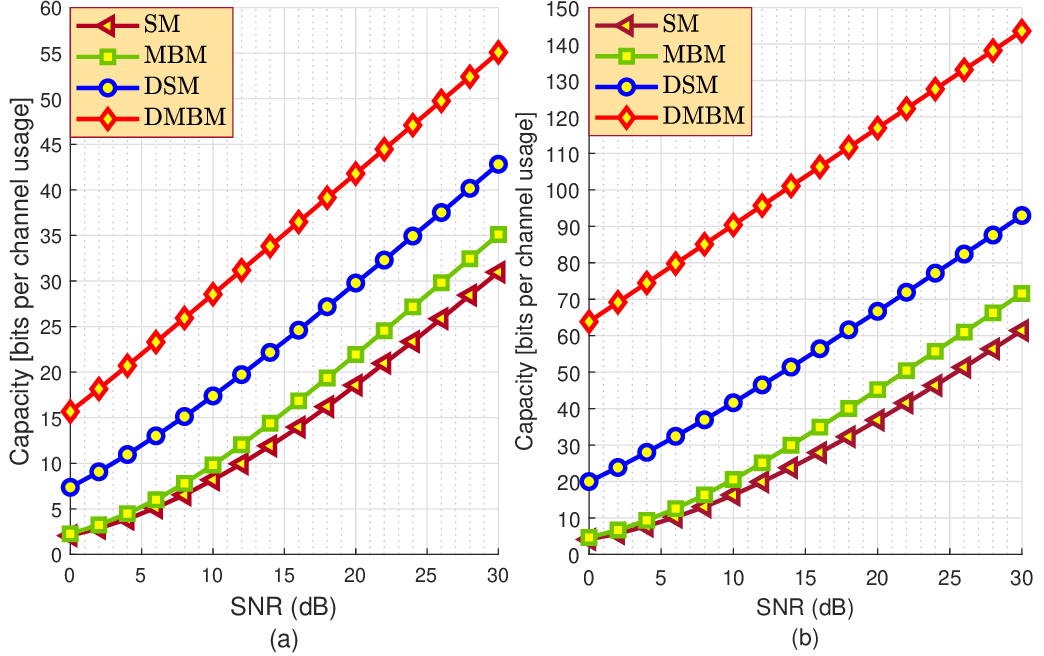}}
	\caption{Capacity comparisons of the SM, MBM, DSM, and DMBM systems, $m_{rf}=4$, $n_T=4$, $n_R=4$ for (a), and $m_{rf}=8$, $n_T=8$, $n_R=8$ for (b).}
	\label{capacity_comp} 
\end{figure}


\begin{figure*}[t]
\centering{\includegraphics[scale=0.5]{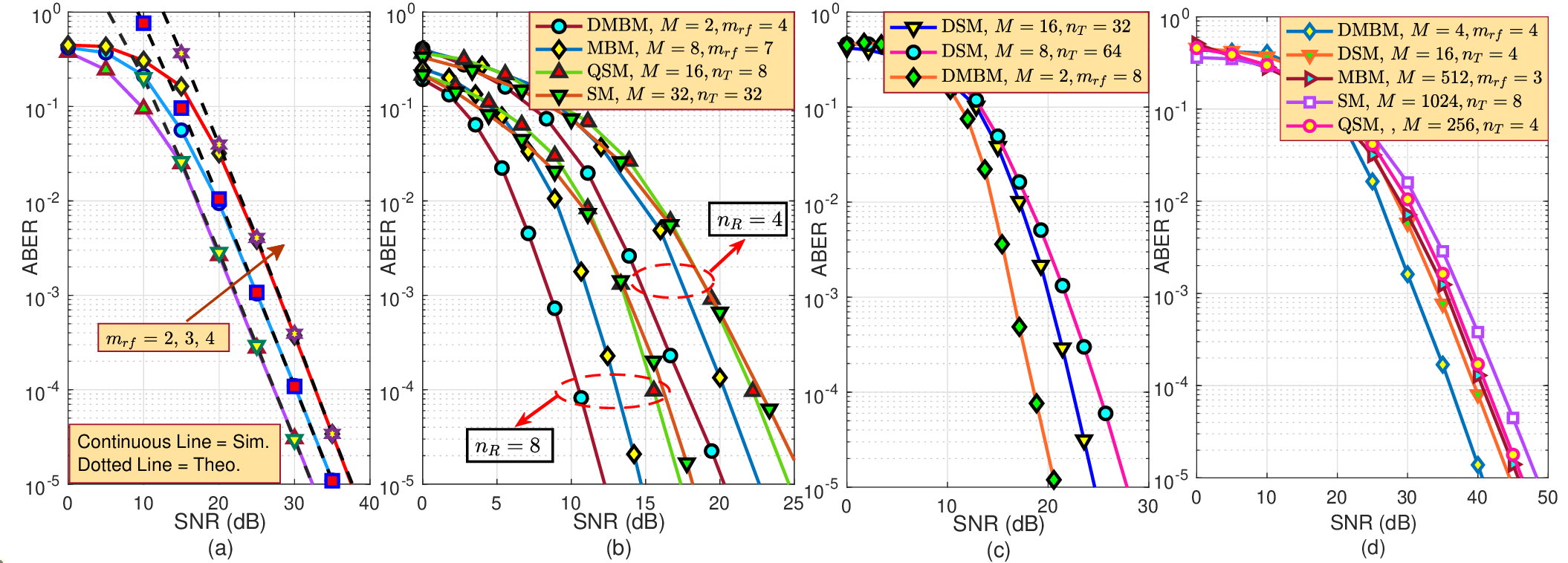}}
	\caption{ (a) Theoretical and simulation performance comparisons of the proposed DMBM system for $M=2$, $n_R=2$ while $m_{rf}=2,3,4$, (b) Performance comparisons of DMBM, MBM, QSM, and SM systems for $n_R=4$ and $n_R=8$ while $\eta=10$, (c)
Performance comparisons of DMBM and DSM systems for $n_R=5$ while $\eta=18$, and (d) 
Performance comparisons of DMBM, DSM, MBM, SM, and QAM systems for $n_R=2$ while $\eta=12$.}
	\label{DMBM_sims_teos} 
\end{figure*}

\section{Capacity Analysis}

In this section, the proposed DMBM system capacity is presented. Capacity analysis will be derived considering the signal constellation and transmission vectors. The ergodic capacity of MIMO systems is defined as follows \cite{capacity}:
\begin{eqnarray} 
\label{capa1}
\mathcal{C} = E_{\textbf{H}} \Big\{ \underset{P_{\textbf{s}_u}}{\text{max}}  \ I\big(\textbf{s}_u;\textbf{y} \: | \: \textbf{H}\big) \Big\},
\end{eqnarray}
where, $\textbf{s}_u \in \mathbb{C}^{2\times 1} =[s_1^u \: (s_2e^{j\phi_{\text{opt}}})^u]^T$, where $u=1,2,\ldots,M^2$, and $\ I\big(\textbf{s}_u;\textbf{y} \: | \: \textbf{H}\big)$ represents the mutual information between the transmission vector $\textbf{s}_u$ and the received vector $\textbf{y}$ for known $\textbf{H}$. However, the major difference between the proposed DMBM system and the traditional MIMO system is that it transmits extra data bits at two active MAP indices. Therefore, the capacity of the proposed DMBM system is the maximization of the mutual information between the transmitted MAP indices and the QAM symbols with $\textbf{y}$  and is defined as follows \cite{anacapa, MIMOFDA}:
\begin{eqnarray} 
\label{capa2}
\mathcal{C} =  E_{\textbf{H}} \Big\{\underset{P_{\textbf{X}_{\omega_1,\omega_2}},P_{\textbf{s}_u}}{\text{max}}  \ I\big(\textbf{X}_{\omega_1,\omega_2}, \textbf{s}_u;\textbf{y} \big) \Big\},
\end{eqnarray}
where $\textbf{X}_{\omega_1,\omega_2}  =[\textbf{x}_1^{\omega_1} \: \textbf{x}_2^{\omega_2}]^T$, here $\textbf{x}_1^{\omega_1} \in \mathbb{C}^{\mathcal{R} \times 1 } = [ 0 \ldots 0  1 0 \ldots 0]^T, \textbf{x}_2^{\omega_2} \in \mathbb{C}^{\mathcal{R} \times 1 }= [ 0 \ldots 0  1 0 \ldots 0]^T$  are vectors with $\textit{k}_1^{\text{th}}$ and $\textit{k}_2^{\text{th}}$ element $1$. The noisy signal expressed in (\ref{received_signal}) is rewritten according to this definition as follows:
\begin{eqnarray} \label{received_signal2} 
\textbf{y} &=& \textbf{H} \textbf{X}_{\omega_1,\omega_2} \textbf{s}_u \ + \ \textbf{w}.
\end{eqnarray} 
In (\ref{capa2}), $I\big(\textbf{X}_{\omega_1,\omega_2}, \textbf{s}_u;\textbf{y} \big)$ is the mutual information, which is the number of bits that can be decoded without errors at the receiver is defined as follows:
\begin{eqnarray} 
\label{capa3}
I\big(\textbf{X}_{\omega_1,\omega_2}, \textbf{s}_u;\textbf{y} \big) = H(\textbf{y}) - H\Big(\textbf{y} \ \big| \ \textbf{X}_{\omega_1,\omega_2}, \textbf{s}_u \Big),
\end{eqnarray}
where $H(.)$ is entropy function.  Considering (\ref{received_signal2}), the entropy of $\textbf{y}$ for the known $\textbf{X}_{\omega_1,\omega_2}$ and $ \textbf{s}_u$ can be expressed as: 
\begin{eqnarray} 
\label{capa4}
H\Big(\textbf{y} \ \big| \ \textbf{X}_{\omega_1,\omega_2}, \textbf{s}_u \Big) = H(\textbf{w}) = \text{log}_2 \big| \pi e N_0 \textbf{I}_{n_R} \big|.
\end{eqnarray} 
where $\textbf{w}\sim\mathcal{CN}(\textbf{0}_{n_S}, N_0\textbf{I}_{n_S})$, $\textbf{y}\sim\mathcal{CN}(\textbf{0}_{n_S}, \sigma^2_\textbf{y}\textbf{I}_{n_S})$ have complex Gaussian distribution, and $e$ is Euler's number. Also, $\textbf{0}_{n_S}$ is an $n_S$ dimensional vector with all elements zero. As shown in (\ref{capa4}), $H\big(\textbf{y} | \textbf{X}_{\omega_1,\omega_2}, \textbf{s}_u \big)$ express the dependence of active MAP indices and QAM symbols. Hence, to maximize the expression $I\big(\textbf{X}_{\omega_1,\omega_2}, \textbf{s}_u;\textbf{y} \big)$, it is necessary to maximize the expression $H(\textbf{y})$. Consequently, the entropy function $H(\textbf{y})$ is obtained as follows:
\begin{eqnarray} 
\label{capa6}
H(\textbf{y}) = -E_\textbf{y} \big\{ \text{log}_2P(\textbf{y})\big\} = \text{log}_2 \big| \pi e \textbf{R}_{\textbf{y}\textbf{y}} \big|,
\end{eqnarray} 
where the expression $\textbf{R}_{\textbf{y}\textbf{y}}$, which is the covariance matrix of the $\textbf{H}\textbf{X}_{\omega_1,\omega_2} \textbf{s}_u$ vector, is expressed as follows:
\begin{eqnarray} 
\label{capa7}
\textbf{R}_{\textbf{y}\textbf{y}} = E(\textbf{y}\textbf{y}^H \big| \textbf{X}_{\omega_1,\omega_2} ) 
=\frac{1}{\mathcal{A} \mathcal{R}} \sum^{\mathcal{R}}_{\omega=1} \!  \textbf{H}\textbf{X}_{\omega_1,\omega_2} \textbf{X}_{\omega_1,\omega_2}^H \textbf{H}^H + N_0 \textbf{I},
\end{eqnarray} 
where $\mathcal{A}$ is the number of columns of $\textbf{X}_{\omega_1,\omega_2}$, which is 2. Finally, using (\ref{capa2})-(\ref{capa7}), the capacity of the proposed DMBM system is derived as follows:
\begin{eqnarray} 
\label{capa8}
\mathcal{C}= E_\textbf{H} \Bigg\{ \text{log}_2 \text{det}\Bigg( \textbf{I}_{n_R} \! + \!\frac{\sum^\mathcal{R}_{\omega_1=1} \sum^\mathcal{R}_{\omega_2=1} \!\textbf{H}\textbf{X}_{\omega_1,\omega_2} \textbf{X}_{\omega_1,\omega_2}^H \textbf{H}^H}{\mathcal{A} \mathcal{R} N_0} \! \Bigg)\! \Bigg\}. 
\end{eqnarray} 

Fig. \ref{capacity_comp} presents the capacity comparisons of the DMBM, SM, MBM, and DSM systems for various system parameters. The system parameters are chosen as $m_{rf}=4$, $n_T=4$, $n_R=4$ for Fig. \ref{capacity_comp} (a), and $m_{rf}=8$, $n_T=8$, $n_R=8$ for Fig. \ref{capacity_comp} (b). Fig. \ref{capacity_comp} clearly shows that the proposed DMBM system has a higher capacity than the SM, MBM, and DSM systems.

\section{Simulation Results of the DMBM System}

In this section, simulation results of the proposed DMBM system are presented for $M$-QAM modulation over Rayleigh fading channels. The BER performances of the SM, QSM, MBM, and DSM systems are provided as a benchmark for comparisons. SNR is defined as $\mathrm{SNR (dB)}=10\log_{10}(E_s/N_0)$, where $E_s$ is the average symbol energy. The optimal ML detector is used for the estimation of the transmitted symbols and active indices at the receiver. All simulation results are obtained under the assumption that fading channels are uncorrelated Rayleigh distributions.

BER vs SNR curves containing the theoretical and simulation results of the proposed DMBM system for different $m_{rf}$ values and $M=2$, $n_R=2$ are given in Fig. \ref{DMBM_sims_teos} (a). The theoretical and simulation curves of the proposed system are obtained quite similar to each other. As it can be understood from Fig. \ref{DMBM_sims_teos} (a), the performance of the DMBM gets better as the $m_{rf}$ value decrease. Also, the continuous and dotted lines represent the simulation and theoretical results, respectively.

In Fig. \ref{DMBM_sims_teos} (b) the BER performance comparisons of the SM, QSM, MBM, DSM, and DMBM systems are presented. According to the results, the DMBM system has been observed to have a better error performance than other compared systems for $\eta=10$. The DMBM system provides $2.36$, $4.36$, and $5.49$ dB for $n_R=4$ and $2.45$, $5.12$, and $5.92$ dB for $n_R=8$ SNR gains according to the MBM, QSM, and, SM systems, respectively.  In addition, as can be seen from Fig. \ref{DMBM_sims_teos} (b), the performance of the systems improves when the number of receive antenna $n_R$ is increased.



The BER performance comparisons of DMBM and DSM systems are shown in Fig. \ref{DMBM_sims_teos} (c) for $\eta=18$. It is seen that the DMBM system provides better performance than the DSM system at the same spectral efficiency. Also, as the spectral efficiency increases, the implementation of the DSM system becomes very difficult because huge amounts of transmitting antennas are needed in the transmitter.

Fig. \ref{DMBM_sims_teos} (d) shows the BER performance of the DMBM system at varying SNR values compared to the other benchmark systems for $\eta=12$. The proposed DMBM system has SNR gains of 3.88, 5.09, 5.63, and 7.13 dB compared to DSM, MBM, QSM, and SM systems, respectively.

As a result, in all comparisons, it is seen that the proposed system gives better performances compared to benchmarks.

\section{Conclusions and Discussions}

This paper presents a novel index modulation technique called DMBM with high spectral efficiency, low energy consumption, and high performance. The proposed system doubles the amount of information sent during a transmission period compared to the traditional MBM method. This significantly increases the spectral efficiency of the DMBM system. Theoretical and simulation analyses are performed for the proposed DMBM and it is observed that the results are consistent with each other. It is shown that the DMBM system provides better performance than other benchmark systems for the same spectral efficiency. Also, the throughput, computational complexity, energy efficiency, spectral efficiency, and capacity analyses of DMBM, DSM, QSM, SM, and MBM systems are obtained and it is shown that the proposed DMBM system is a better wireless data transmission system with an acceptable complexity.


\bibliographystyle{IEEEtran}
\bibliography{IEEEabrv,Referanslar}

\end{document}